\documentclass{iaus}
\usepackage{graphicx}

\title[Formation of binaries] %% give here short title %%
{Angular momentum of two collided rarefied preplanetesimals and
formation of binaries}

\author[Sergei I. Ipatov]   %% give here short author list %%
{Sergei I. Ipatov$^{1,2}$}
\affiliation{$^1$Catholic University of America \\ Washington DC, USA 
\\ email: {\tt siipatov@hotmail.com} \\[\affilskip]
$^2$Space Research Institute, Moscow, Russia}

\pubyear{2010}
\volume{263}  %% insert here IAU Symposium No.
\pagerange{37--40}
\jname{Icy Bodies of the Solar System}
\editors{J.A. Fernandez, D. Lazzaro, D. Prialnik, \& R. Schulz, eds.}
\begin{document}

\maketitle

\begin{abstract}
The mean angular momentum associated with the collision of two celestial objects 
moving in almost circular heliocentric orbits was studied. 
The results of these studies were used to develop models of the formation of binaries 
at the stage of rarefied preplanetesimals.
The models can explain a greater fraction of binaries 
formed at greater distances from the Sun.
Sometimes there could be two centers of contraction inside the rotating 
preplanetesimal formed as the result of a collision between two rarefied preplanetesimals. 
Such formation of binaries 
could result in binaries with almost the same masses of components separated by a large distance. 
%In particular, the model could be useful for understanding the formation of
%binaries with close masses that 
%are separated by a large distance and that have any values of eccentricity and 
%inclination of the mutual orbit of the secondary and the primary components. 
Formation of a disk around the primary
%some other binaries 
could result because the angular momentum that was 
obtained by a rarefied preplanetesimal formed by collision was greater than the 
critical angular momentum for a solid body. 
%During the contraction of such a rotating
%rarefied preplanetesimal, some material could form a disk of material moving around the primary. 
One or several
satellites of the primary %(moving mainly in low eccentric orbits) 
could be formed from the disk.
%% originated during the contraction of the preplanetesimal.
%this cloud at any distance from the primary inside the Hill sphere, but the typical 
%distance would be much smaller than the radius of the sphere. 
%For any trans-Neptunian 
%binary that has been discovered, the angular momentum is smaller than the typical angular 
%momentum of two identical rarefied preplanetesimals having the same total mass as the 
%discovered binary and encountering up to the Hill sphere from circular heliocentric orbits.
\end{abstract}

\firstsection % if your document starts with a section,
\section{Introduction}
%\section{Analysis of images}

In recent years, new arguments in favor of the model of rarefied preplanetesimals -- 
clumps were made (e.g., Cuzzi et al. 2008, Johansen et al. 2007, Lyra et al 2008). 
Even before new arguments in favor of formation of planetesimals 
from rarefied preplanetesimals were developed, Ipatov (2001, 2004) considered that some trans-Neptunian 
objects (TNOs), planetesimals, and asteroids with diameter $d$$>$100 km could be formed directly by 
the compression of large rarefied preplanetesimals, but not by the accretion of smaller solid 
planetesimals. Some smaller objects (TNOs, planetesimals, asteroids) could be debris from larger 
objects, and other smaller objects could be formed directly by compression of preplanetesimals. 
There are several hypotheses of formation of binaries for a model of solid bodies 
(e.g., Petit et al. 2008, Richardson \& Walsh 2006, Walsh et al. 2008). 
Ipatov (2004) supposed that a considerable fraction of trans-Neptunian binaries could be formed at the 
stage of compression of rarefied preplanetesimals moving in almost circular orbits. 
%The models of binary formation due to the gravitational interactions 
%or collisions of future binary components with an object (or objects) that were 
%inside their Hill sphere, which were studied by several authors for solid objects, 
%could be more effective for rarefied preplanetesimals. 
%%For example, due to the almost 
%circular heliocentric orbits of preplanetesimals, the duration of the motion of 
%preplanetesimals inside the Hill sphere could be longer, and the minimum distance 
%between centres of masses of preplanetesimals could be smaller than for solid bodies, 
%many of which could have greater eccentricities of heliocentric orbits.
Based on analysis of the angular momentum of two collided rarefied preplanetesimals,
Ipatov (2009a-b) studied models of 
the formation of binaries at the stage of %rarefied 
the preplanetesimals.

\section{Angular momentum of two collided rarefied preplanetesimals}  

Previous papers devoted to the 
formation of axial rotation of forming objects considered mainly a model of solid-body accumulation. Besides 
such model, Ipatov (1981a-b, 2000, 2009b) also studied the formation of axial rotation for a model of rarefied preplanetesimals. 
He presented the formulas for the angular momentum of two collided rarefied preplanetesimals -- Hill spheres 
(with radii $r_1$ and $r_2$ and masses $m_1$ and $m_2$)  moved in circular heliocentric orbits. 
At a difference in their semimajor axes $a$ equaled to $\Theta(r_1+r_2)$, 
%the tangential velocity of collision is vτ=kΘ∙(G∙MS)1/2∙(r1+r2)∙a-3/2 and 
the angular momentum is 
$K_s=k_{\Theta}(G \cdot M_S)^{1/2}(r_1+r_2)^2m_1m_2(m_1+m_2)^{-1}a^{-3/2}$, 
where $G$ is the gravitational constant, 
and $M_S$ is the mass of the Sun. At $r_a=(r_1+r_2)/a\ll \Theta$,
% and $r_a\ll 1$, 
one can obtain $k_{\Theta} \approx (1-1.5\Theta^2)$. 
The mean value of $k_{\Theta}$ equals to 0.6. 
Mean positive values of $k_{\Theta}$ and mean negative values of $k_{\Theta}$
are equal to 2/3 and -0.24, respectively. 
The values of $K_s$ are positive at $0<\Theta<0.8165$ 
%$0<\Theta<(2/3)^{1/2}\approx0.8165$ 
and are negative at $0.8165<\Theta<1$. 
%The minimum value of $k_{\Theta}$ equals -0.5. 
%In the case of uniform 
%distribution of $\Theta$, the probability to get a reverse rotation is about 1/5.

For homogeneous spheres at $k_{\Theta}$=0.6, $a$=1 AU, and $m_1$=$m_2$, 
the period of axial rotation $T_s\approx9\cdot10^3$ 
hours for the rarefied preplanetesimal formed as a result of the collision of two preplanetesimals -- 
Hill spheres, and $T_s\approx0.5$ h for the planetesimal of density $\rho$=1 g cm$^{-3}$ formed from the preplanetesimal. For greater $a$, the values of 
$T_s$ are smaller (are proportional to $a^{-1/2}$). Such small periods of axial rotations cannot exist, 
especially if we consider bodies obtained by contraction of rotating rarefied preplanetesimals, 
which can lose material easier than solid bodies. 
%The value of $v_s=2\pi r_f/T_s$ (the velocity of a 
%particle on a surface of a rotating spherical object of radius $r_f$ at the equator) is equal to 
%vcf=(G∙mf/rf)1/2 (the minimum velocity at which a particle can leave the surface) at period 
%Tsc=Ts=(3π)1/2∙(ρG)-1/2. 
For $\rho$=1 g cm$^{-3}$, the %value of the %vs 
 velocity of a particle on a surface of a rotating spherical object 
%of radius $r_f$ 
at the equator
is equal to the circular and the escape velocities at 3.3 and 2.3 h, respectively.
%$T_{se}=T_s=(3\pi)^{1/2}(2\rhoG)^{-1/2}$. Tsc≈3.3h and Tse≈2.33h. 
%
%The form of present small bodies can differ much from 
%their primordial form, and their rotation could change due to collisions with solid small bodies.

For five binaries, the angular momentum $K_{scm}$ of the present primary and secondary components 
(with diameters $d_p$ and $d_s$ and masses $m_p$ and $m_s$), the momentum $K_{s06ps}$ of two collided preplanetesimals 
with masses 
of the binary components moved in circular heliocentric orbits at $k_{\Theta}$=0.6, and the momentum 
$K_{s06eq}$ of two identical collided preplanetesimals with masses equal to a half of the total 
mass of the binary components at $k_{\Theta}$=0.6 are presented in the Table. All these three momenta 
are considered relative to the center of mass of the system. $K_{spin}$ is the spin momentum of 
the primary. $L$ is the distance between the primary and the secondary, 
$r_H$ is the radius of the Hill sphere, and $T_{sp}$ is the period 
of spin rotation of the primary. 
%The used data were taken from Wikipedia. 

\begin{table}
%\begin{minipage}{160mm}
\caption{Angular momenta of several binaries.} \label{anymode}
\begin{tabular}{@{}lllllll}
\hline
binary & Pluto & (90842) Orcus  & 2000 CF$_{105}$ & 2001 QW$_{322}$ &  (90) Antiope \\
\hline
$a$, AU          & 39.48  & 39.3  &  43.8  & 43.94  &  3.156 \\
$d_p$, km        & 2340   & 950   &  170   & 108?   &   88    \\
$d_s$, km        & 1212   & 260   &  120   & 108?   &   84    \\
$m_p$, kg  & $1.3\times10^{22}$ & $7.5\times10^{20}$ & $2.6\times10^{18} $ ?&
 $6.5\times10^{17}$ ?   &   $4.5\times10^{17}$    \\
$m_s$, kg  & $1.52\times10^{21}$ & $1.4\times10^{19}$  & 
$9\times10^{17} $ ?& $6.5\times10^{17}$ ?  &   $3.8\times10^{17}$    \\
  &   &   for $\rho$=1.5  & &  &    \\
$L$, km        & 19,750 & 8700 & 23,000  & 120,000 & 171 \\
$L/r_{H}$  & 0.0025   & 0.0029  &  0.04  & 0.3   &   0.007 \\
%$2L/d_p$  & 16.9  & 18.3  &  271  & 2200   &   3.9 \\
$T_{sp}$, h  & 153.3   & 10  &     &    &   16.5 \\
$K_{scm}$, kg~km$^2$~s$^{-1}$ & $6\times10^{24}$ & $9\times10^{21}$ &
$5\times10^{19}$ & $3.3\times10^{19}$ &  $6.4\times10^{17}$ \\
$K_{spin}$, kg~km$^2$~s$^{-1}$ & $10^{23}$ & $10^{22}$ &
$1.6\times10^{18}$ & $2\times10^{17}$ &   $3.6\times10^{16}$ \\
 & & & at $T_s$=8 h& at $T_s$=8 h &   \\
$K_{s06ps}$, kg~km$^2$~s$^{-1}$ & $8.4\times10^{25}$ & $9\times10^{22}$ &
$1.5\times10^{20}$ & $5.2\times10^{19}$ &   $6.6\times10^{18}$ \\
$K_{s06eq}$, kg~km$^2$~s$^{-1}$ & $2.8\times10^{26}$ & $2\times10^{24}$ &
$2.7\times10^{20}$ & $5.2\times10^{19}$ &  $6.6\times10^{18}$ \\
$(K_{scm}+K_{spin})/K_{s06ps}$  & 0.07   & 0.2  &  0.3 & 0.63   &   0.1 \\
$(K_{scm}+K_{spin})/K_{s06eq}$  & 0.02   & 0.01  &  0.2 & 0.63   &   0.1 \\
%$v_{\tau eq06}$, m~s$^{-1}$ & 6.1   & 2.2  &  0.36 & 0.26   &   0.82 \\
%$v_{\tau pr06}$, m~s$^{-1}$ & 5.5   & 1.8  &  0.3 & 0.26   &   0.82 \\
%$v_{esc-pr}$, m~s$^{-1}$ & 15.0   & 5.8  &  0.8 & 0.53   &   1.7 \\
\hline
\end{tabular}
%\end{minipage}
\end{table}

\section{Models of formation of binaries}
%Comparison of angular momenta of present binaries with model angular momenta} 

For circular 
heliocentric orbits, two objects that entered inside the Hill sphere could move there for a longer 
time than those entered the sphere from eccentric heliocentric orbits. The diameters of preplanetesimals 
were greater than the diameters of solid planetesimals of the same masses. Therefore, the models of 
binary formation due to the gravitational interactions or collisions of future binary components 
with an object (or objects) that were inside their Hill sphere, which were studied by several 
authors for solid objects, could be more effective for rarefied preplanetesimals.

We suppose that formation of some binaries could be caused by that the angular momentum that 
they obtained at the stage of rarefied preplanetesimals was greater than that could exist for 
solid bodies. During contraction of a rotating rarefied preplanetesimal, some material with velocity greater than the circular velocity
%$vs>vcf$ 
could have formed a cloud (that transformed into a disk) of material that moved around the 
primary. One or several satellites of the primary could be formed from this 
cloud. Some material could leave the Hill sphere of a rotating contracting planetesimal, 
and the mass of an initial rotating preplanetesimal could exceed the mass of a corresponding 
present binary system. Due to tidal interactions, the distance between binary components 
could increase with time, and their spin rotation could become slower. For the discussed 
model of formation of binaries, the vector of the original spin momentum of the primary 
was approximately perpendicular to the plane where the secondary component (and all other 
satellites of the primary) moved. It is not necessary that this plane was close to the ecliptic 
if the difference between the distances from centers of masses of collided preplanetesimals 
to the middle plane of the disk of preplanetesimals was comparable with sizes of 
preplanetesimals. Eccentricities of orbits of satellites of the primary formed in such 
a way are usually small.
As it was shown by Ipatov (2009b), the 
critical angular momentum could be attained as a result of a collision of two identical asteroids of any radii ($<$6000 km). 
At the same eccentricities of heliocentric orbits and $m_1/m_2$=const, the probability to attain the 
critical momentum at a collision is greater for smaller values of $m_1$ 
($m_1 \ge m_2$) and $a$. 

Some collided rarefied preplanetesimals had a greater density at distances closer to their centers. 
It might be possible that sometimes there were two centers of contraction inside the 
rotating preplanetesimal formed as a result of a collision of two rarefied preplanetesimals. 
Such formation of binaries 
could result in binaries with almost the same masses of components separated by a large distance. 
%In such cases, the values 
%of the eccentricity of the orbit of the secondary component can be different.
 It could be also 
possible that the primary had partly contracted when a smaller object (objects) entered into the 
Hill sphere, and then the object was captured due to collisions with the material of the outer 
part of the contracted primary. For such a scenario, a satellite can be formed at any distance (inside 
the Hill sphere) from the primary.
The eccentricity 
of the mutual orbit of components can be any (small or large)
for the model of two centers of contraction.

For the binaries 
presented in the Table, the ratio $r_K=(K_{scm}+K_{spin})/K_{s06eq}$ is smaller than 1. 
%For most of 
%observed binaries, this ratio is smaller than for the trans-Neptunian binaries considered. 
Small values of $r_K$ for most discovered binaries can be due to that preplanetesimals already had been
partly compressed at the moment of collision.
%, i.e. they were smaller than their Hill 
%spheres and/or could be denser for distances closer to the center of a preplanetesimal.  

%For an object of density $\rho$, 
At $K_s$=const, $T_s$ is proportional to $a^{-1/2}\rho^{-2/3}$. 
Therefore, for greater $a$, more material of a contracting rotating preplanetesimal was 
not able to contract into a primary and could form a cloud surrounding the 
primary (or there were more chances that there were two centers of contraction). 
This can explain why binaries are more frequent among TNOs than among large main-belt 
asteroids, and why the typical mass ratio of the secondary to the primary is greater 
for TNOs than for asteroids. Longer time of contraction of rotating preplanetesimals 
at greater $a$ (for dust condensations, this was shown by several authors, e.g. by Safronov) 
could also testify in favor of the above conclusion. 
%Most of rarefied preasteroids could 
%turn into solid asteroids before they collided with other preasteroids. 
Spin and form of 
an object could change during evolution of the Solar System.
% due to collisions with other solid small bodies. 

%The axial tilt of Pluto is 119.6$^\circ$. Reverse rotation does not contradict to a collision of two objects 
%(see the previous section). Present rotation of Sylvia around 
%its short axis is not in favor of our model, and it could be obtained at a stage of solid bodies.

%Solid bodies could get critical angular momenta (corresponding to $T_s$$<$3.3 h) at collisions in the 
%case of eccentric heliocentric orbits. Ipatov (2009b) showed that 
%$T_s\approx 6.33r_f/v_\tau$ at a collision of two 
%identical bodies, where $r_f$ is the radius of the formed body and $v_\tau$
%is the tangential component of the collision velocity.
%%$T_s$ is proportional to $m_1/m_2$  at $m_1\gg m_2$) and homogeneous spheres. 
%At $v_\tau$=3.5 km s$^{-1}$, the equality $6.33r_f/v_\tau$=3.3 h is fulfilled at $r_f\approx6600$ km. Therefore, 

Ipatov (2009b) discussed the possibility of a merger of two 
rarefied preplanetesimals
and the formation of highly elongated %promordial 
small bodies by the merger of two (or several) partly compressed components.
%Some collided solid bodies 
%with the critical momentum could be disrupted due to collision. 
\section{Conclusions}

Some trans-Neptunian objects could have acquired their primordial axial momenta and/or satellites 
at the stage when they were rarefied preplanetesimals. Most rarefied 
preasteroids could have become solid asteroids before they collided with other preasteroids. 
Some collided rarefied preplanetesimals could have greater densities at 
locations that are closer to their centers. In this case, there sometimes could 
be two centers of contraction inside the rotating preplanetesimal formed 
as a result of the collision of two rarefied preplanetesimals. Such 
contraction could result in binaries with similar masses separated by 
any distance inside the Hill sphere and with any value of the 
eccentricity of the orbit of the secondary component relative to 
the primary component. The observed separation 
distance can characterize the radius of a greater encountered preplanetesimal.

The formation of some binaries could have resulted because the angular 
momentum of a binary that was obtained at the stage of rarefied preplanetesimals 
was greater than the angular momentum that can exist for solid bodies.  
Material that left a contracted preplanetesimal 
formed as a result of a collision of two preplanetesimals  
could form a disk around the primary.
One or more satellites of the primary could be grown in the disk
at any distance from the primary inside the Hill sphere,
but typical separation distance is much smaller than the radius of the sphere.
The satellites moved mainly in low eccentric orbits. 
Both of the above scenarios could have taken place at the same time. In this case, 
it is possible that, besides massive primary and secondary components, 
smaller satellites could be moving around the primary and/or the secondary.

For discovered trans-Neptunian binaries, the angular momentum is usually considerably
smaller  than the typical angular momentum of 
two identical rarefied preplanetesimals having the same total mass 
and encountering up to the Hill sphere from circular heliocentric orbits. 
This conclusion is also true for preplanetesimals with masses of components 
of considered trans-Neptunian binaries. The above difference in momenta 
and the separation distances, which usually are much smaller 
than the radii of Hill spheres, support the hypothesis that most preplanetesimals 
already had been partly compressed at the moment of collision, 
i.e. were smaller 
than their Hill spheres and/or were 
denser at distances closer to the center of a preplanetesimal. 
The contraction of preplanetesimals could be slower farther
 from the Sun, which can explain the greater
fraction of binaries formed at greater distances from the Sun.

%The results of our studies of angular momentum of encountered objects 
%can be used for analysis of the formation of axial 
%rotation of rarefied preplanetesimals and solid planetesimals.

\end{document}